\documentclass[a4paper,aps,prb,reprint,showpacs]{revtex4-1}
\usepackage{amsmath,amssymb,mathrsfs}
\usepackage[pdftex]{graphicx}

\begin{document}

\bibliographystyle{unsrt}

\title{Spin-orbit coupling and proximity effects in metallic carbon nanotubes.}

\author{Piotr Chudzinski}
\affiliation{Institute for Theoretical Physics, Center for Extreme Matter and Emergent Phenomena, Utrecht University, Leuvenlaan 4, 3584 CE Utrecht, The Netherlands}

\date{\today}

\begin{abstract}

We study the spin-orbit coupling in metallic carbon nanotubes (CNTs)
within the many-body Tomonaga-Luttinger liquid (TLL) framework.
For a well defined sub-class of metallic CNTs, that contains both
achiral zig-zag as well as a sub-set of chiral tubes, an effective
low energy field theory description is derived. We aim to describe
systems at finite dopings, but close to the charge neutrality point
(commensurability). A new regime is identified where the spin-orbit
coupling leads to an inverted hierarchy of mini-gaps of bosonic
modes. We then add a proximity coupling to a superconducting (SC)
substrate and show that the only order parameter that is supported
within the novel, spin-orbit induced phase is a topologically
trivial s-SC.

\end{abstract}

\maketitle

\section{Introduction}

In the past few years, we have witnessed a renewed interest in superconducting (SC) proximity
effects in 1D systems. The reason why this topic is in the forefront
of condensed matter research was the discovery\cite{Kane-topoSC,
Fujimoto-topoSC2} that a SC with a topologically non-trivial order
parameter is able to support the long sought Majorana surface
states\cite{Wilczek-viewpoint-Majorana}. Moreover, it was
shown\cite{Alicea-topSC-prox} that the nontrivial SC can be
artificially created by a proximity coupling of a trivial
superconductor with a 1D wire that has a substantial spin-orbit
coupling. While the first experimental
signatures\cite{Kouwenhoven-the-Maj-experim} that such a device can
indeed support Majoranas fuelled the interest of the community, at the
same time questions about the role of disorder\cite{Atland-disord,
Lee-disord}, low dimensionality breaking and electron-electron
interactions\cite{Braunecker-interac} were raised. To avoid at
least the first two issues one may consider a carbon nanotube
(CNT), a self organized, strictly 1D system that nowadays can be
produced with ultra-clean quality. However there is still an issue
of interactions and moreover one could wonder if the peculiar
spin-orbit coupling, that is present in CNT, can produce
topologically non-trivial proximity effect. The answer to this
questions turns out to be negative and this is one of the main results of this paper.

The price of moving from a simple wire, with e.g. cubic structure,
to a CNT is that, in the latter case, one deals with a highly
non-trivial mapping between real and reciprocal space structures.
The low energy physics of a nanotube can be derived from that of a
hexagonal graphene lattice by imposing a quantization condition
along the CNT circumference. For concreteness, we consider a CNT
with a chiral vector $(n,m)$ such that $(n-m)~mod~3 = 0$. Then,
within the sub-bands that follow from circumferential momentum
quantization, there exist a sub-band which falls very close to the
Dirac points $K,K'$ of a graphene reciprocal space. The nanotube
is metallic and the vicinities of the two distinct Dirac points
are called valleys. More refined analysis includes a curvature
induced shift\cite{Kane+Mele-gaps} away from Dirac points
$\Delta_{curv}$ as well as a spin-orbit
coupling\cite{ando:jpsj2000} that, in the sublattice basis, have
both diagonal $\Delta'_{SO}$ and non-diagonal $\Delta_{SO}$
components\cite{Jeong-curvature-SO}. The spin-orbit coupling is a
subject of particular interest due to its peculiar nature, with
larger non-diagonal $\Delta_{SO}$ component.

It is tempting to incorporate the spin-orbit couplings (and
$\Delta_{curv}$) on a single-particle level because then their
only effect is to change the band structure. So far, all
attempts\cite{egger:prb2012, sau:prb2013} to address non-trivial
proximity effects in CNT were based on such single particle
framework. However, neglecting the electron-electron interactions $V(q)$ would
have been justified only if they were a tiny perturbation added on
the top of $\Delta_{SO}$ and $\Delta_{curv}$. In reality:
$V(q\approx 0)\sim 0.3eV$ and $V(q\approx 2|K|)\sim 10meV$
\cite{KaneBalents-armchair, Egger-CNT-TLL-big} while
$\Delta_{curv}\approx\Delta_{SO}\leq 1 meV$
\cite{Jeong-curvature-SO, ando:jpsj2000, kuemmeth:nature2008,
jespersen:prl2011} so that one faces exactly the opposite hierarchy of
energy scales. Also, at a more fundamental level, a key property of
1D systems is that even upon introducing an infinitesimally small
$V(q)$, their low energy description must be given in terms of
collective excitations\cite{Desphande-Nature-rev}. A carbon
nanotube (CNT) is no exception from this general principle. A well
established fact is that the velocity of charge fluctuations is
strongly renormalized\cite{KaneBalents-armchair,
Egger-CNT-TLL-big}. This is one manifestation of strong
correlations in the physics of CNTs and it implies that a naive
refermionization back to the original electrons' framework is not
allowed.

It is then an important task to incorporate the effects of
spin-orbit coupling into a proper many body description of CNT. To
this end a few partial problems have already been solved. In
Ref.\onlinecite{Egger-diagonal-spin-orbit}, under an assumption
that there exists a mini-gap in the single particle spectrum, it
has been shown that the diagonal component $\Delta'_{SO}$ is able
to shift velocities and TLL parameters of all TLL modes. This
shift can be understood (see discussion of
Eq.\ref{eq:ham-TLL-def2}) if one remembers that the on-site
component is uniform in space, thus it has a density-density form.
Furthermore, a detailed analysis of $\Delta_{curv}$ term (and
interaction induced terms of the same form) done for a zig-zag
tube, exactly at half filling, was done in
Ref.\onlinecite{Carr-CNT-dimerization}. A crucial assumption was
that the system is deep inside a Mott insulating phase. The aim our work is to go beyond this special case and study a new
physics generated by the $\Delta_{so}$ away from commensurability.

A further novelty is that a sub-set of chiral tubes is also covered. Apart from
extending the range of validity, this also erases any
constraints between $\Delta_{curv}$ and $\Delta_{so}$. For instance
$\Delta_{curv}$ can be varied by a tube's
twist\cite{Eggert-gap-twist} (not possible for achiral CNTs) or,
due to absence of a lattice inversion symmetry, an unprotected
$\Delta_{so}$ can be
modified by higher order scattering processes. This versatility
allows us to freely tune the parameters of our model.

The paper is organized as follows. In Sec.\ref{sec:model} we identify a class of chiral tubes where our theory applies and then term by term we introduce a description within the 1D framework. The following section Sec.\ref{sec:RG} aims to derive an effective low energy description in a renormalization group (RG) spirit. It is divided in two parts, a high energy part Sec.\ref{ssec:high-en}, which is dominated by holon behaviour, while in Sec.\ref{ssec:low-en} dedicated to the lower energies we use adiabatic approximation and focus on gap opening in the spin/valley modes. Then, in Sec.\ref{sec:proxim}, we check the influence of spectral gaps on superconducting proximity effects. Finally, in Sec.\ref{sec:discus}, we discuss an issue of experimental detection of the gaps, an influence of the other symmetry breaking terms, like e.g. valley-mixing term, and other SC orderings proposed for nanotubes. The paper is closed with conclusions, Sec.\ref{sec:concl}, and two appendices that contains estimates for a holon expectation value and for a proximity hybridization with a substrate.


\section{CNT as a two leg ladder}\label{sec:model}

The hamiltonian of a CNT can be written as:
\begin{equation}\label{eq:H-gen}
  H=H_{0}+H_x+H_{di}+H_{so}
\end{equation}
where $H_{0}$ is a TLL hamiltonian (see Eq.\ref{eq:ham-TLL-def2}), $H_x$ (Eq.\ref{eq:Hx-intro}) contains the many-body interactions with large momentum exchange, $H_{di}$ (Eq.\ref{eq:H-dim}) is a dimerization potential introduced to capture $\Delta_{curv}$, in a way following Ref.\onlinecite{Carr-CNT-dimerization}, and the last, new term $H_{so}$ (Eq.\ref{eq:H-so}) contains the $\Delta_{SO}$.

\begin{figure}
  \centering
  \includegraphics[width=\columnwidth]{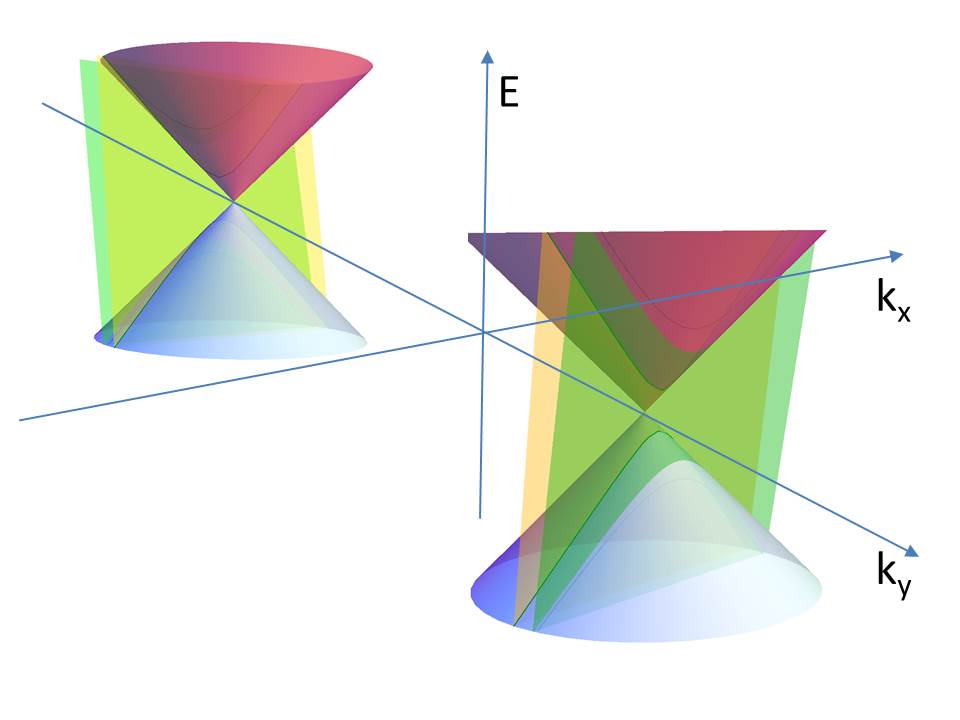}\\
  \caption{Low energy band structure of a CNT that falls within the zig-zag (like) class. As pointed out in Ref.\onlinecite{KaneBalents-armchair}, due to anisotropy of $t_{\vec{r},\vec{d}}^{\bar{\sigma}}$ the cones are slightly shifted away from $K, K'$ points of a reciprocal unit cell. The plane cross sections are due to circumferential quantization condition, only those values of quantized $k_y$ that are closest to Dirac points are shown. On each cone there are two dispersions $E(k_x)$. The mean shift away from Dirac point opens a gap $\Delta_{curv}$, while a split between the two dispersions is proportional to $\Delta_{so}$. Small tilt of the planes is proportional to $\Delta'_{so}$, in Ref.\onlinecite{Jeong-curvature-SO} details concerning these spin-orbit effects are given.}\label{fig:bands}
\end{figure}

The real space hamiltonian in fermionic language reads:
\begin{equation}\label{eq:ham-ferm-real}
 H_{\bar{0}+di+so}= \sum_{\vec{r}}\sum_{\vec{d}} (t_{\vec{r},\vec{d}}^{\bar{\sigma}} a_{\vec{r},\bar{\sigma}}^{\dag}b_{\vec{r}+\vec{d},\bar{\sigma}}+h.c)
\end{equation}
where we have taken a nearest neighbour hopping on a bipartite lattice. The $a_{\vec{r}}^{\dag}$ operator creates an electron on the lattice site A with coordinates $\vec{r}$. The summations go over all lattice sites positions $\vec{r}$ and all nearest neighbours $\vec{d}$, thus $\vec{d}$ is a linear combination of hexagonal lattice vectors. Due to curvature effects the hopping parameter, a complex number $t_{\vec{r},\vec{d}}^{\bar{\sigma}}$, is anisotropic in $\vec{r}$-space and spin dependent. On the top of Eq.\ref{eq:ham-ferm-real} one adds an electron-electron interaction which has a Coulomb character. In order to extract the low energy physics one turns to the reciprocal space description with a momenta $k_x,k_y$ directed along CNT's axis and circumference respectively. For a chiral tube both these axis make a finite angle with a helical line along which the graphene lattice is folded. The resulting band structure is illustrated in Fig.\ref{fig:bands}. The two cones, that are characteristic of hexagonal lattice, are cut in slices that stem from the circumferential quantization set on $k_y$.

We put a chemical potential close enough to the Dirac points such
that in the following we can restrict ourself only to the lowest
lying sub-bands in each of the two valleys. Creation operator $c^{\dag}_{k,\bar{\sigma}\alpha}$ are assigned to these states, where $\bar{\sigma}$ is a spin index, an index $\alpha=K,K'$ and
$k$ is a component of an electron momentum along a CNT, thus 1D physics is
implicitly assumed. Then two Fermi points are present near each
Dirac point and this leads to a system with overall four Fermi
points. It must be the two leg ladder model that describes the low
energy physics for this band structure. An exact mapping between
real space and $c_k^{\dag}$ has been found for achiral
armchair\cite{KaneBalents-armchair} and zig-zag
tubes\cite{Carr-CNT-dimerization}. We take a closer look at the later
ones as these can accommodate finite $\Delta_{curv}$ and
$\Delta_{so}$, the subject of this study. The zig-zag CNT is
mapped onto a ladder with an interchain $t_{\perp}=0$ and this
allows us to identify chains (of an abstract ladder) with valleys (of
graphene).

The validity of this simple mapping can be extended also onto a
sub-class of chiral tubes. In a recent work\cite{moj-2classCNT} we have
showed that it is possible to distinguish a class of tubes defined
by a condition $(n-m)/gdc(m,n)~mod~3\neq 0$, that have two pairs of Fermi
points located around $K_{\perp}\neq 0,~K_{||}\approx0$, that is similar to the
zig-zag CNT. In Ref.\onlinecite{moj-2classCNT} we considered an
infinitely sharp, local chemical potential, an extra term in the hamiltonian $\sim \mu_0
\delta{x-x_0} \rho(x)$ with $\mu_0 \rightarrow \infty$ and
$\rho(x)$ is an electronic density, a Fourier transform of $\sum_k
c_k^{\dag}c_{k+q}$. For the zig-zag like tubes a response to such
potential is a reflection matrix that is strictly diagonal in the valley space.
From this it follows that an operation
$c^{\dag}(x=x_0)|\Psi_k\rangle$ (where $|\Psi_k\rangle$ is an
eigenstate) is diagonal in the valley space. We apply creation
operation infinitely many times along a CNT to find that $\int dx~
c^{\dag}(x)|\Psi_k\rangle $ is also diagonal in valley space which
implies that a valley$\equiv$chain description, with
$t_{\perp}=0$, should be valid for these chiral tubes, at least in
the regime close to the Dirac points ($k_{||}\approx 0$) which is
of interest in this study. To quantify the criterion, by
analogy with commensurate-incommensurate transition\cite{Tsuchiizu-commen-incommen}, we
notice a competition between the inter-chain interaction terms
$\sim g_{ic} \cos\phi_{\rho-}$ (see below for definition of bosonic
fields and Eq.\ref{eq:Hx-intro}) and the inter-chain hopping (present for $k_{||}\neq 0$) that upon
bozonization gives a term $\sim t_{\perp}\cos\theta_{\rho-}$.
These bosonic expressions, that contain two canonically conjugated fields, suggest that the following criterion for $k_{||}$ can be given
$t\sin(k_{||}a)< g_1$. Substituting numerical values, this implies
that our reasoning can be safely applied when the doping $\delta<0.03$.
The fact that a CNT can be described as valley$\equiv$chain ladder
is enough to apply to the results of this work.

We go directly to the bosonization description of the lowest
sub-band fermions. We follow a standard procedure. First, one
extracts the long wavelength behaviour around the Fermi points:

$$\bar{\psi}_{\bar{\sigma}\alpha}(x)=\exp(i k_F
x)\psi_{R\bar{\sigma}\alpha}(x)+\exp(-i k_F
x)\psi_{L\bar{\sigma}\alpha}(x)$$

where we have written the formula in terms of a real space field $\bar{\psi}_{\bar{\sigma}\alpha}(x)$ which is an eigenvalue of the of the second quantization
operators $c_{\bar{\sigma}\alpha}(x)$ (Fourier transform of $c^{\dag}_{k,\bar{\sigma}\alpha}$), in the Fock space of the coherent states. Then
one focuses on the slow components of the fluctuations around Fermi points and introduces the
collective bosonic fields:

\begin{equation}\label{eq:fermi-bose-def}
\psi_{R,L\bar{\sigma}\alpha}(x)=\kappa_{R,L\bar{\sigma}\alpha}\frac{1}{2\pi
a}\exp(i[\phi_{\bar{\sigma}\alpha}(x)\pm\theta_{\bar{\sigma}\alpha}(x)])
\end{equation}

where $\kappa_{R,L\bar{\sigma}\alpha}$, the Klein factors, ensure
proper anti-commutation relations. The collective fields can be
also expressed directly through the real-space density operator
defined before (for the valley-diagonalization argument), for
instance
$\phi_{\bar{\sigma}\alpha}(x)=-\pi\nabla\rho_{\bar{\sigma}\alpha}(x)$.
Finally one turns to a total/transverse basis by a transformation
$$
\phi_{\rho\pm}=\frac{1}{2}(\phi_{1\uparrow}+\phi_{1\downarrow}-\phi_{2\uparrow}-\phi_{2\downarrow});~\phi_{\sigma\pm}=\frac{1}{2}(\phi_{1\uparrow}+\phi_{1\downarrow}-
\phi_{2\uparrow}-\phi_{2\downarrow})$$

Four collective modes $\phi_{\nu}$ (and canonically conjugate
$\theta_{\nu}$) are present: total/transverse charge/spin modes
($\nu=\rho\pm,\sigma\pm$). The total charge mode $\rho+$ is sometimes called a holon as it contains an electric charge of a hole, while the other three modes are neutral and contain the spin/valley component. With these bosonic modes defined, we
can now write down each part of Eq.\ref{eq:H-gen}. The $H_{0}$
reads:

\begin{multline}\label{eq:ham-TLL-def2}
    H_{0}[\phi_{\nu}]= \sum_{\nu} \int \frac{dx}{2\pi}
    \left[(v_{\nu}K_{\nu})(\partial_{x} \theta_{\nu})^{2}+(\frac{v_{\nu}}{K_{\nu}})(\partial_{x} \phi_{\nu})^{2}\right]
\end{multline}

The main advantage of working in the bosonization framework is that an
entire $V(q\approx 0)$ part of interactions is already included in
Eq.\ref{eq:ham-TLL-def2}. Since in CNTs  the interactions have a long
range Coulomb character, the small momentum exchange interactions are much larger than those with large momentum exchange. The Coulomb interactions bosonize as:

$$ H_{Cou}=\frac{2e^2}{\pi}\int dx \int dx'
V(r-r')\partial_x \phi_{\rho+}(x)\partial_{x'} \phi_{\rho+}(x')$$

Clearly, only the total charge mode (holon) is affected. Because of this, the holons' velocity $v_{\rho+}$ can be up
to five times larger than $V_F$, while $K_{\rho+}$ can be as small
as 0.25. Velocities of all the other so called neutral modes, remain
$\approx V_F$.

The large momentum exchange part of electron-electron interactions, where $V(q\approx |K|)$, or in other words the part that cannot be written in a density-density form that is $\neq \int dx \int dx'\rho(x)\rho(x')$, adds several non-linear terms:
\begin{widetext}
\begin{multline}\label{eq:Hx-intro}
    H_x = \frac{1}{2(\pi a)^2}\int dr g_3
    \cos(2\phi_{\rho+}-2\delta x)(\cos(2\phi_{\rho-})+\cos(2\phi_{\sigma+})+\cos(2\phi_{\sigma-})+\cos(2\theta_{\sigma-}))-g_{1c}\cos(2\phi_{\rho-})\cos(2\phi_{\sigma+})\\
    -g_{2c}\cos(2\phi_{\rho-})\cos(2\phi_{\sigma-})+g_{1a}\cos(2\phi_{\sigma+})\cos(2\theta_{\sigma-})+g_{\parallel c}\cos(2\phi_{\rho-})\cos(2\theta_{\sigma-})+g_{1}\cos(2\phi_{\sigma+})\cos(2\phi_{\sigma-}))
\end{multline}
\end{widetext}

where the backscattering terms, with spin and/or valley index
change in the process, are indicated as $g_{1,2i}$. We use
notation from Ref.\onlinecite{moj-ladder} and convention for the
Klein factors as in Ref.\onlinecite{moj-ladder,
Egger-CNT-TLL-big}. The only difference is that the
Ref.\onlinecite{moj-ladder} is dedicated to two-leg ladders with
large $t_{\perp}$ (more customary case) while here $t_{\perp}=0$
but a finite inter-chain interaction $V_{\perp}$ is present. To
transfer between the two models it is enough to make an
interchange $\cos2\phi_{\rho-}\Leftrightarrow \cos2\theta_{\rho-}$
in Eq.\ref{eq:Hx-intro}. Terms $\sim g_3$ in Eq.\ref{eq:Hx-intro}
are umklapp scattering terms which transfer two left movers into
right movers (or \emph{vice-versa}). This requires
commensurability with the lattice, obeyed at half filling, while
at finite doping $\delta$ these are gradually suppressed.

Additional terms, dimerization and spin-orbit coupling, are present because the $C_3$ symmetry of the underlying
graphene lattice is broken upon wrapping. A $\sigma^*-\pi^*$ hybridization, induced by wrapping, changes the hopping amplitude along the tube circumference and this shifts the position of the Dirac points\cite{KaneBalents-armchair}. The lowest energy sub-bands are defined independently by the quantization condition along the tube circumference, so they are now shifted with respect to the new Dirac cones. This effectively results in an opening of a so-called \emph{mini-gap} in the spectrum, the $\Delta_{curv}$. As it was proven in Ref.\onlinecite{Carr-CNT-dimerization} this effect can be grasped by introducing a dimerization potential into the effective 1D hamiltonian of CNT, Eq.\ref{eq:H-gen}. Such a term, the Peierls term, is well known in 1D systems, it is exactly solvable via Bogoliubov transformation in the particle-hole channel and leads to a gap opening in the single particle spectrum. In bosonization language it reads\cite{Carr-CNT-dimerization}:
\begin{multline}\label{eq:H-dim}
  H_{di}=\frac{1}{2\pi a}\int dx g_d(-\cos(\phi_{\rho+}-\delta x )\cos\phi_{\rho-}\cos\phi_{\sigma+}\cos\phi_{\sigma-}\\
  -\sin(\phi_{\rho+}-\delta x )\sin\phi_{\rho-}\sin\phi_{\sigma+}\sin\phi_{\sigma-})
\end{multline}
where $g_d=V_{di}/V_F$ generates mini-gaps $\Delta_{curv}$ at $K,K'$
points. From the Bogoliubov transformation, done for an alternating
potential in a single-particle limit, we know that in the lowest
order the relation is simple $V_{di}=\Delta_{curv}$. However
$V_{di}$ incorporates also further terms, the staggered potential
terms that are produced in the course of the RG
flow\cite{Carr-CNT-dimerization}. The sole term $V_{di}$,
Eq.\ref{eq:H-dim}, written in bosonization language, contains a sort
of "frustration": there is a competition between terms perfectly compensating each other. Sines and
cosines wish to lock $\phi_{\nu}$ fields at different
minima. When $V_{di}$ dominates the physics, then the bosonic
framework is inappropriate, instead one should turn back to the
original fermions (to obtain the Peierls transition). But this simple prescription does not
work if there are other terms, like electron-electron interactions, present as well. Then it is
necessary to write $H_{di}$ (and $H_{so}$) in the bosonization
language in order to take advantage of the adiabatic
approximation\cite{Nersesyan-CNT-orders, Carr-CNT-dimerization,
Levitov-narrow-gaps-TLL} and separate out the influence of the fast
$\phi_{\rho+}$ field. In Ref.\onlinecite{Carr-CNT-dimerization} the
"frustration" problem was solved by considering a regime dominated
by the umklapp scattering (deep inside the Mott phase) which
favors $\cos\phi_{\rho+}$ and then also other cosines
automatically follow. Below we show a different mechanism that is able to lift the "frustration".

The spin-orbit coupling shifts band dispersions away from the Dirac points by an amount that depends on the spin/valley degree of freedom of a fermion, in an opposite direction for electrons with opposite helicities\cite{kuemmeth:nature2008}. Alternatively, this phenomenon can be seen as a spin-dependent variation of a mini-gap in the spectrum around the point where bonding and anti-bonding bands used to cross in the tight-binding model. As a result, in the single particle picture, $\Delta_{so}$ adds a spin/valley dependent component to the mini-gaps, see Fig.\ref{fig:bands}. By reasoning along the same lines like for the curvature term, this can be interpreted as an extra spin-valley dependent single-particle backscattering. The $\Delta_{so}$ term is then expected to have a form similar
to Eq.\ref{eq:H-dim}, with the only difference that the left/right
mixing term now involves the z-Pauli matrices in spin and valley
spaces\cite{egger:prb2012}:
$$\hat{O}_{so}\sim
c_{LK\uparrow}^{\dag}c_{RK\uparrow}-c_{LK'\uparrow}^{\dag}c_{RK'\uparrow}-c_{LK\downarrow}^{\dag}c_{RK\downarrow}+
c_{LK'\downarrow}^{\dag}c_{RK'\downarrow}$$
The spin-orbit coupling is expressed in the spin-valley
basis because of the intricate topological origin of the effect\cite{Jeong-curvature-SO, ando:jpsj2000, kuemmeth:nature2008,
jespersen:prl2011}: electrons of opposite valleys are precessing along the helical lines of opposite twist. However we have established
that, within our effective two leg ladder description, the valley degree of freedom can be associated with the chain
index. Then, in Eq.\ref{eq:fermi-bose-def}, $\alpha=K,K'$ and thanks to that $\hat{O}_{so}$ has a simple bosonized expression. Finally, the spin-orbit term that is off-diagonal in the sub-lattice space, asks to choose a bond (not an on-site) operator to be hermitian. These few constraints are enough to deduce the following form of spin-orbit term $\Delta_{so}$ in the bosonic language:
\begin{multline}\label{eq:H-so}
  H_{so}=\frac{1}{2\pi a}\int dx g_{so}(-\cos(\phi_{\rho+}-\delta x)\cos\phi_{\rho-}\cos\phi_{\sigma+}\cos\phi_{\sigma-}\\
  +\sin(\phi_{\rho+}-\delta x)\sin\phi_{\rho-}\sin\phi_{\sigma+}\sin\phi_{\sigma-})
\end{multline}

One immediately notices that thanks to an opposite sign of the two terms in Eq.\ref{eq:H-so},
the $g_{so}$ is able to lift the "frustration" present in the sole $H_{di}$.

\section{RG treatment of cosine terms}\label{sec:RG}

As usual in the RG procedure, we inspect how the parameters of the hamiltonian are effectively changing upon integrating out high-energy degrees of freedom. The RG flow is divided in two
stages: the first when the doping is negligible and the system
flows like if it was at commensurate filling, the second when doping is significant and only
the backscattering terms in Eq.\ref{eq:Hx-intro} should be kept.

\subsection{High energy RG flow}\label{ssec:high-en}

The first stage of RG flow stops at energy scale $\Lambda'$ that is defined by the condition $\delta[\Lambda']=1$. Above this energy RG is dominated by the umklapp and dimerization/spin-orbit terms whose perturbative, single loop, RG equations read:

\begin{align}\label{eq:RG-system}
\dot{g}_3&=3g_3(1-K_{\rho+})\\
\dot{g}_{d,so}&=g_{d,so}(2-(K_{\rho+}+K_{\rho-}+K_{\sigma+}+K_{\sigma-})/4)
\end{align}

where, in the first equation, we used the fact that $K_{\rho-}\approx K_{\sigma+}\approx K_{\sigma-}\approx 1$, otherwise three different equations for three different umklapp channels would need to be given. The reason why Eq.\ref{eq:RG-system} dominate is because in CNT, in the UV limit, a relevant parameter range is $0.2<K_{\rho+}\ll 1$ thus one can safely assume $|K_{\rho+}-1|\gg|K_{\nu\neq\rho+}-1|$ and then all terms that
contain the $\phi_{\rho+}$ mode are much more relevant than others. The umklapp has a scaling dimension $d_{3}=1-K_{\rho+}$ while
the $g_d$ and $g_{so}$ are even more relevant with $d_d=1.25-K_{\rho+}/4$.

The RG flow of other non-linear terms is determined by the following equations:
\begin{align}\label{eq:RG-system2}
\dot{g}_{1c}&=g_{1c}(2-(K_{\rho-}+K_{\sigma+}))\\
\dot{g}_{2c}&=g_{2c}(2-(K_{\rho-}+K_{\sigma-}))\\
\dot{g}_{1a}&=g_{1a}(2-(K_{\sigma-}^{-1}+K_{\sigma+}))\\
\dot{g}_{\parallel c}&=g_{\parallel c}(2-(K_{\sigma-}^{-1}+K_{\rho-}))\\
\dot{g}_{1}&=g_{1}(2-(K_{\sigma-}+K_{\sigma+}))\\
\end{align}
While this flow is much slower in the first stage of RG, in the second stage of RG  Eq.\ref{eq:RG-system2} becomes the driving force.

The TLL parameters are also renormalized:
\begin{widetext}
\begin{align}\label{eq:RG-system3}
\dot{K}_{\rho+}&=-\frac{1}{2}K_{\rho+}^2(4g_3^2+g_d^2+g_{so}^2)J_0(\delta)\\
\dot{K}_{\rho-}&=-\frac{1}{2}K_{\rho-}^2(J_0(\delta)(g_3^2 +g_d^2+g_{so}^2)+g_{1c}^2+g_{2c}^2++g_{\parallel c}^2)\\
\dot{K}_{\sigma+}&=-\frac{1}{2}K_{\sigma+}^2(J_0(\delta)(g_3^2+g_d^2+g_{so}^2)+g_{1c}^2+g_{1a}^2+g_1^2)\\
\dot{K}_{\sigma-}&=-\frac{1}{2}K_{\sigma-}^2(J_0(\delta)(g_d^2+g_{so}^2)+g_1^2+g_{2c}^2)+g_{1a}^2+g_{\parallel c}^2\\
\end{align}
where $J_0(\delta)$ is a Bessel function of the first kind (we take UV cut-off equal to one).

\end{widetext}

The bare (initial) amplitudes of the exchange terms in
Eq.\ref{eq:RG-system2} are small but finite and were thoroughly
calculated in Ref.\onlinecite{Egger-CNT-TLL-big}. In that language:
$g_{1c}=g_1=f$, $g_{2c}=b-f$ and $g_{1a}=g_{\parallel c}=b$, where
$b,f$ are amplitudes of large momentum scattering processes
computed on a microscopic CNT lattice for armchair tube. The estimate $b,f\approx (0.05,0.1)V(q=0)\approx (0.005,0.01)V_F$ was given and in our chiral case we are likely to be close to the upper limit since in a less symmetric lattice certain cancellation between real space Coulomb interactions are not exact. On the top of it, in our
non-armchair case, there is a contribution from a coupling between
orbital momenta of two electrons. It enhances
$g_{1c},g_{1},g_{\parallel c}$ (a ferro-orbital configuration of
initial orbital momenta $\mu_o$ implies that the two carriers will
repel each other) and reduces $g_{2c},g_{1a}$ (an
antiferro-orbital configuration of initial orbital momenta
$\mu_o$). In CNTs $\mu_o$ can be an order of magnitude
larger\cite{minot:nature2004} than $\mu_B$ which makes this unusual contribution to
electron-electron interactions worth considering. To estimate it we can
compare it with $\Delta_{so}\approx |\mu_o||\mu_B| \leq 1meV$. $\Delta_{so}$ originates from similar mechanism, an interaction
between $\mu_o$ and $\mu_B$ as a carrier moves along a helical
line of a CNT.
The umklapp terms correspond to terms with even larger momentum
exchange, thus their initial (UV) amplitudes are smaller for the Coulomb-like interactions.
Moreover their amplitude is further suppressed by a finite doping and this suppression is two times faster than for the $V_{di}$ amplitude.

Our study is dedicated to the case of a finite doping. Since in the later part of the
paper the SC proximity effects are considered, we must take a
model with a non-zero conductance on an interface with a
substrate, thus a model with a constant chemical potential. Then the
doping is not a constant but a renormalizable quantity that competes with
interactions. This effect we incorporate in the following RG equation:
\begin{equation}\label{eq:RG-system4}
\dot{\delta}=\delta-(3g_3^2+g_{d}^2+g_{so}^2)J_1(\delta)
\end{equation}

\begin{figure}
  \includegraphics[width=\columnwidth]{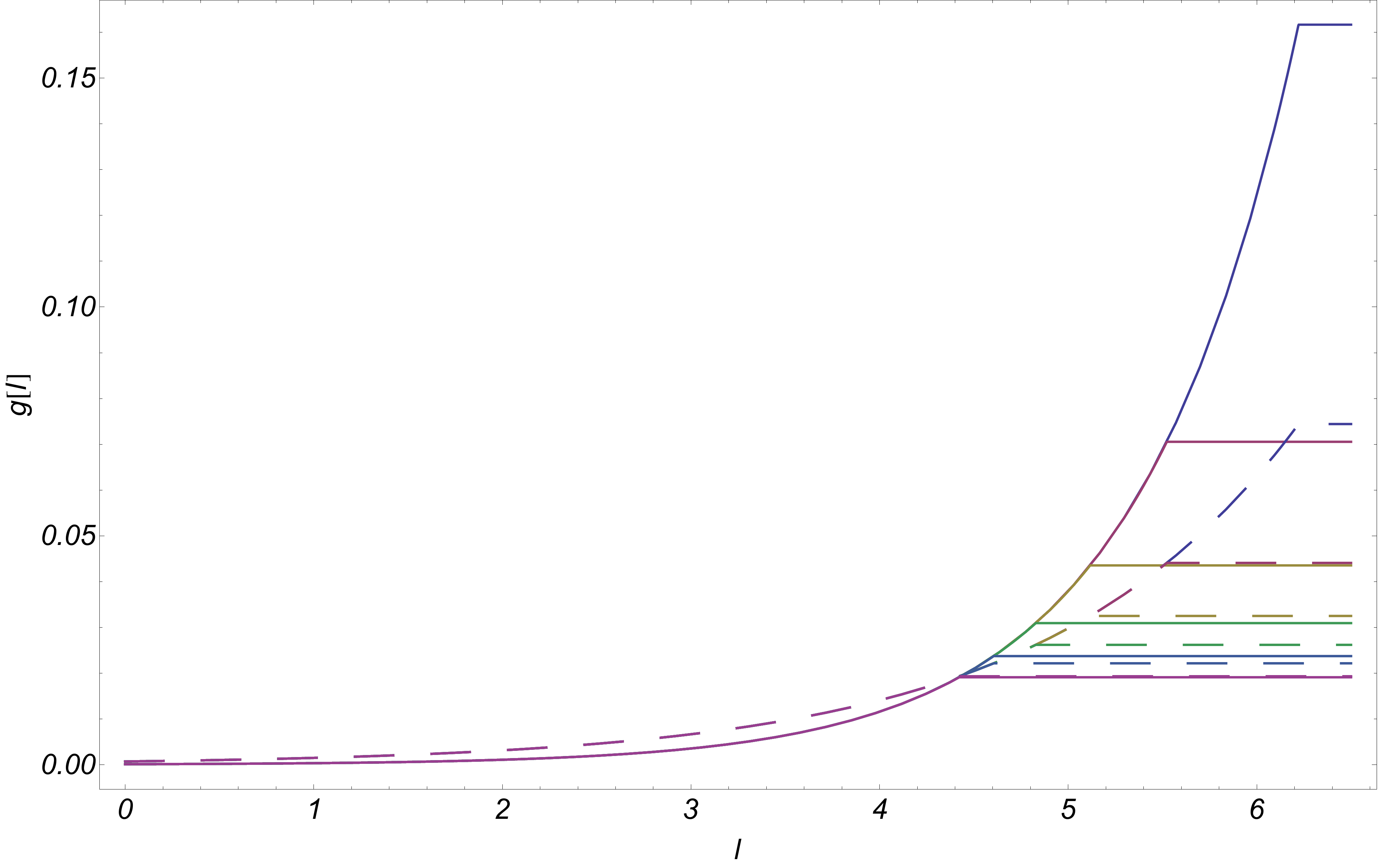}\\
  \caption{The RG flow of couplings that are violently relevant in the first stage of RG: the umklap (dashed lines) and dimerization (solid lines) terms. Different doping levels are shown by different color code (from top to bottom): from $\delta=0.002$ to $\delta=0.012$ with intervals $0.002$. One notices that in a point when flow stops, thus for $l=\Lambda'$ the dimerization term is always stronger at finite dopings even though we started with $g_{d}=g_{so}=0.0001$ and $g_3=0.0007$ (and $K_{\rho+}=0.25$).}\label{fig:RG-flow}
\end{figure}

where $J_1(\delta)$ is a Bessel function of the first kind. The RG flow of $\delta$ (in the first stage of RG) can produce two
outcomes: i) $\delta[l]$ rapidly grows and when $\delta[l]\sim
1$ then this first stage of RG must be stopped and $g_3$ terms in
Eq.\ref{eq:Hx-intro} and $g_d$ (and $g_{so}$) terms Eq.\ref{eq:H-dim}-\ref{eq:H-so}
effectively drop out of the problem because the integrands in Eq.\ref{eq:Hx-intro}-\ref{eq:H-so} contain rapidly oscillating terms; ii) $\delta[l]$ rapidly drops
to zero then the system flows to a Mott or Peierls physics, where a competition
between $g_{3}$ and $g_{di}$ (and $g_{so}$) determines the low energy properties. Case ii) can be realized only when $\delta[\Lambda]<g_i^2$ which for CNTs translates into extremely small doping levels. Nevertheless, for a finite $\delta[l]$ during RG, this competition persists and since the dimerization is less affected by doping then this phase should
expand. Crucially, as we show later in Sec.\ref{ssec:anti-adiab}, the nature of the "dimerized" phase changes when
the energy scale $\Lambda'$ is of the same order or smaller than $\Delta_{so}$.

Close to commensurate filling, for a parameters range that is relevant for a CNT, we identify quite a broad regime where
$g_d$ (and $g_{so}$) dominate over $g_3$ terms. We analyze several RG flows for initial parameters: $g_{d}=g_{so}=0.0001$,
$g_3\in (0.0001,0.001)$, $K_{\rho+}\in (0.25,0.35)$ (these values are relevant for CNTs) and $\delta[l=0]\in
(0.001,0.012)$. Some examples of
RG flows for different $\delta$, are given in Fig.\ref{fig:RG-flow}. We observe that both
terms grows and, in a chosen range of parameters, the
dimerization term is always the dominant one, even if one starts
with (an overestimated) ratio $g_3[\Lambda]/g_d[\Lambda]\simeq 10$. The flow stops
for $l_1 \in (5.5, 7)$ which taking initial UV cut-off
$\Lambda=1.5eV$ translates into an energy scale $\Lambda' \sim
10^{-3}eV$ that is comparable with the bare $\Delta_{so}$. The values reached by
$g_3$ and $g_d$ (and $g_{so}$) at $\Lambda'$ are substantial $g_3
< g_d\sim 10^{-1}$ (see Fig.\ref{fig:RG-flow}) but still below
$\sim 10^0$, thus gaps are not open yet. While these terms drops out of
RG but in the following should be considered as a substantial perturbation.

\subsection{Physics at energies below $\Lambda'\sim 1meV$}\label{ssec:low-en}

\subsubsection{Anti-adiabatic approximation}\label{ssec:anti-adiab}

We restrict ourself to $H=H_0+H_{di}+H_{so}$. At $\Lambda'$ we re-analyze the theory using the adiabatic approximation\cite{Nersesyan-CNT-orders}. To be precise we use an anti-adiabatic version of it to focus on the physics of three neutral modes.
Following Ref.\onlinecite{Levitov-narrow-gaps-TLL} we separate out the
fast $\phi_{\rho+}$ field using an auxiliary variable
$\eta(x)=\arctan[(\sin\phi_{\rho-}\sin\phi_{\sigma+}\sin\phi_{\sigma-})/(\cos\phi_{\rho-}\cos\phi_{\sigma+}\cos\phi_{\sigma-})]$.
After shifting the field $\tilde{\phi}_{\rho+}(x)\rightarrow
(\phi_{\rho+}(x)+\delta x)+\eta(x)$ the action is separable. Than for the fast field we obtain a sine-Gordon model:
\begin{equation}\label{eq:ham-fast}
  H_{\phi_{\rho+}}=H_{0}[\phi_{\rho+}]-\int dr M[\phi_{i\neq \rho+}] \cos(\tilde{\phi}_{\rho+}(x))
\end{equation}
where the mass term

$$M[\phi_{i\neq\rho+}]=(V_{so}(l)+V_{di}(l))|_{\Lambda'}\sqrt{1+\sum_{\nu\neq\iota}\cos
2\phi_{\nu}\cos2\phi_{\iota}}$$

can be obtained using identities:
$\arctan(\sin(\alpha/\beta))=\alpha/(\alpha^2+\beta^2)^{-1}$ and
$(\sin\phi_{\rho-}\sin\phi_{\sigma+}\sin\phi_{\sigma-})^2+(\cos\phi_{\rho-}\cos\phi_{\sigma+}\cos\phi_{\sigma-})^2=1+\sum_{j\neq
i}\cos 2\phi_j \cos2\phi_i$. While writing Eq.\ref{eq:ham-fast} we
neglect terms $\sim\eta(x)$ (and higher powers) and derivatives
$\sim\partial_t \eta(x)$, which is justified in the adiabatic
limit (slow $\eta(x)$) and in the presence of substantial $V_{so}$ (then $\eta(x)\rightarrow 0$ is justified). The $V_{so}$, as written in
Eq.\ref{eq:H-so}, favors cosines' over sines' minima and thus provided
$V_{so}\sim \Lambda'$ we tend to a well defined limit
$\eta\rightarrow 0$, variations of $\eta$ field are gradually
suppressed.

\subsubsection{Effective hamiltonian for the slow fields}

For the slower fields we proceed by integrating out the
$\tilde{\phi}_{\rho+}$. At energies $\sim \Lambda'$ the
Eq.\ref{eq:ham-fast} is a sine-Gordon model thus
$\langle\cos(\tilde{\phi}_{\rho+}(x))\rangle|_{\omega=\Lambda'}\neq
0$ (see Appendix for details). Then, upon
expanding $M[\phi_{i\neq \rho+}]$ we arrive at an emergent
non-linear term:

\begin{multline}\label{eq:action-after-integrating}
    H_{\tilde{di}}[\phi_{i\neq \rho+}]=-g'_d(\cos(2\phi_{\rho-})\cos(2\phi_{\sigma+})\\
    +\cos(2\phi_{\sigma-})[\cos(2\phi_{\sigma+})+\cos(2\phi_{\rho-})])
\end{multline}

where $\tilde{g}_d\sim g_d \langle\cos(\tilde{\phi}_{\rho+}(x))\rangle$
and in the lowest approximation the expectation value is
proportional to the symmetry breaking term
$\langle\cos(\tilde{\phi}_{\rho+}(x))\rangle \sim V_{so}$. In the
sign convention we use both hamiltonians Eq.\ref{eq:H-dim} and
Eq.\ref{eq:action-after-integrating} are minimized by the same
combination of locked neutral fields, thus validating our mapping.
The Eq.\ref{eq:action-after-integrating} should be combined with
the backscattering part of $H_x$(Eq.\ref{eq:Hx-intro}).
The following perturbation to $H_{0}[\phi_{\nu\neq\rho+}]$ emerges:
\begin{multline}
H_{\tilde{x}}= -\tilde{g}_{1c}\cos(2\phi_{\rho-})\cos(2\phi_{\sigma+})+\\
    \cos(2\phi_{\sigma-})[-\tilde{g}_{2c}\cos(2\phi_{\rho-})+\tilde{g}_{1}\cos(2\phi_{\sigma+})]+\\
    \cos(2\theta_{\sigma-})[\tilde{g}_{1a}\cos(2\phi_{\sigma+})+\tilde{g}_{\parallel c}\cos(2\phi_{\rho-})]
\end{multline}

The initial parameters for 2nd stage of RG flow (we take a new UV
cut-off $\Lambda'$) are determined by the values obtained in the
end of the 1st stage.  The RG flow of $H_x$ is a BKT flow with the parameters that falls
close to the negative separatrix\cite{Egger-CNT-TLL-big} (the SU(2) invariant line on the
$g-K$ plane, with the RG flowing straight away from the critical
point). Then, in the first RG stage,
$g[l]/V_F=g[\Lambda]/V_F/(1-g[\Lambda]l)$ and one finds that
$g_{1c}[\Lambda'],g_{2c}[\Lambda'],g_{1}[\Lambda']\sim 10^{-2}$.
However, some of the
$g'_i[\Lambda']$ terms are larger because they contain also
$g_d[\Lambda']$ contribution (which is significant even when multiplied by
$\langle\cos\tilde{\phi}_{\rho+}\rangle|_{\omega=\Lambda'}$). To be precise:
$\tilde{g}_{1c}[\Lambda']=g_{1c}[\Lambda']+\tilde{g}_d[\Lambda']$,
$\tilde{g}_{2c}[\Lambda']=g_{2c}[\Lambda']+\tilde{g}_d[\Lambda']$,
$\tilde{g}_{1}[\Lambda']=g_{1}[\Lambda']-\tilde{g}_d[\Lambda']$ while all other terms
are not affected. As for the
TLL parameters, in the first part of the RG flow the umklapp,
dimerization and spin-orbit terms all involve
$\sim\cos(\phi_{\rho-})$. As a result the RG flow changes the TLL
parameter $K_{\rho-}$ downwards (already initially, at
$l=\Lambda_0$, this term is shifted slightly below $K_{\rho-}=1$
by interactions\cite{Egger-CNT-TLL-big} as well as
$\Delta'_{SO}$\cite{Egger-diagonal-spin-orbit}), same holds for
$K_{\sigma+}$. Thus we conclude that for
$g_{1c}\cos(\phi_{\sigma+})\cos(\phi_{\rho-})$ term we make a shift
upwards along the negative separatrix, that is both $g_{1c}$ and
$1-K_{\rho-}$ change upwards. Since close to
the separatrix the gap $\Delta=\Lambda \exp (-V_F/\tilde{g}_{1c})$, the
dependence is exponential. Taking quite a conservative estimate that
both $g_{1c}[\Lambda']$ and $\tilde{g}_d[\Lambda']$ are of the same order
we find that the exponent is reduced by a factor of two in comparison with
Ref.\onlinecite{Egger-CNT-TLL-big}. This leads to much enhanced
gaps $\Delta_{\rho-,\sigma+}\sim 0.1meV$. Numerical (RG) calculations
confirm this finding: $g'_{1c}[l']=1$ already for
$l'\approx 2$. A gap opens in the spectrum of the two bosonic
modes $\phi_{\sigma+},\phi_{\rho-}$. The fields are locked at an energy minima $\phi_{\sigma+}=0,\phi_{\rho-}=0$. The gap value is equal to the mass
of a soliton of the sine-Gordon model
$M_{\rho-,\sigma+}=2\sqrt{2\tilde{g}_{1c}u_{\rho-}/\pi K_{\rho-}}$, from
this $M_{\rho-,\sigma+}\approx 0.1 meV$. The two
estimates coincide.

The RG flow of $\sigma-$ mode is more difficult to follow. In
$H_x$ (Eq.\ref{eq:Hx-intro}) we find competing
$\cos(\phi_{\sigma-})$ and $\cos(\theta_{\sigma-})$ terms which exactly compensate each other, also in the lowest energy sector when some modes acquire gaps. Moreover, this
implies that $dK_{\sigma-}/dl\approx 0$, while to begin with
$K_{\sigma-}=1$ and even accounting for the diagonal spin-orbit
coupling\cite{Egger-diagonal-spin-orbit}, the $\Delta'_{SO}$, does
not move $K_{\sigma-}$ from the marginal value $K_{\sigma-}=1$. Thus we conclude that this mode is in a self-dual point, at least within the manifold of interaction terms we decided to take into account. Usually, such a situation is treated by employing re-fermionization\cite{gogol-tsvel_book_1d}, then separating real/imaginary parts as Majorana fermions, e.g. $\xi_{0,i}=Re[\exp\phi_{\sigma-}(x_i)]$, and finally using fusion rules to map the hamiltonian onto a doublet of quantum Ising chains $H_0[\sigma-]+H_{\tilde{x}}[\sigma-]=\sum_{l=0,1}\sum_i \sigma^z_{l,i}\sigma^z_{l,i}+h\sigma^{x}_{l,i}$ with order/disorder operators $\sigma_{0,1},\mu_{0,1}$ defined as $\xi_{0,i}=\sigma_{0,i}\mu_{0,i}$. In Ref.'s\onlinecite{Egger-CNT-TLL-big,
Nersesyan-CNT-orders} the procedure was used in the context of CNTs.  The self dual point is equivalent, in the Ising model language, to $\sigma_1$ chain passing through criticality. The other Ising chain is always gapped and, by accounting for a negative sign of the mass term, we deduce that the order Ising operators $\sigma_{0}$ have a finite amplitude, which means that $\sin\phi=0$ and $\sin\theta=0$ while both respective cosines are non-zero. This does not allow us to identify the unique ground state, but only to narrow down the possibilities. Since $K_{\rho+}<1$ it shall be DW ordering, either intra-valley  CDW or intra-valley SDWz, with either bond or on-site character. One must remember that there are other ordering possibilities e.g. squared order parameters, with higher periodicities, which may be dominant when $K_{\rho+}<0.25$.  Furthermore, since self-duality is not protected by any symmetry, one cannot exclude that due to some extremely tiny perturbation, not accounted in our generic model, a gap in $\phi_{\sigma-}$ actually opens. However this  depends on the finer details of a CNT under consideration and describe physics that takes place at energies $\sim 10^{-9}eV$ or below\cite{Egger-CNT-TLL-big}, so we refrain from its further analysis here.

Even though the exact ground state remains elusive, the larger gaps $M_{\rho-},M_{\sigma+}$ that certainly open, provide sufficient conditions to determine the allowed proximity effect.

\section{Proximity effects}\label{sec:proxim}

The inverted hierarchy of gaps plays an important
role in the proximity effect. This is because usually the coupling with the substrate and the superconducting gap (on the surface) are smaller than $M_{\rho-,\sigma+}$.  In the appendix we give a brief description of the hybridization, in the fermionic language. To understand how these microscopic considerations are linked with many body TLL theory, one must sum over
all sites of the CNT within a unit cell, turn to collective fields and then express the result in the two-leg ladder
basis. This is a well established procedure, we follow
Ref.\onlinecite{Nersesyan-CNT-orders} to find that the singlet SC
order operators in a zig-zag (like) CNT are:
\begin{equation}\label{eq:s-SCdef}
    \hat{O}_{s}^{SC}\sim\exp(\imath\theta_{\rho+})(\cos\phi_{\rho-}\cos\phi_{\sigma+}\cos\theta_{\sigma-}
    +\imath(\sin\leftrightarrow\cos))
\end{equation}
\begin{equation}\label{eq:m-SCdef}
    \hat{O}_{m}^{SC}\sim\exp(\imath\theta_{\rho+})(\cos\phi_{\rho-}\sin\phi_{\sigma+}\sin\theta_{\sigma-}+
    \imath(\sin\leftrightarrow\cos))
\end{equation}

The first one, $\hat{O}_{s}^{SC}$, corresponds to a purely local tunneling process (a pair is created on one site) and thus it is more likely to occur in type-II superconductor (short coherence length), than $\hat{O}_{m}^{SC}$ when a pair is created non-locally (with different phase on adjacent sites). Both $\hat{O}_{s,m}^{SC}$ have a scaling dimension
$d_{\Delta}=2-(3+K_{\rho+}^{-1})/4$, thus they are relevant for
$K_{\rho+}>0.2$. This holds when we assume $K_{\nu}=1$ for $\nu \neq
\rho+$, accounting for the fact that actually (in the low energy limit) $K_{\rho-}<1$ changes the condition to $K_{\rho+}>0.25$. It is likely that the condition $K_{\rho+}>0.25$ is fulfilled when a CNT lies on
a conducting substrate which provides the screening for Coulomb interactions and thus reduce their range. The relevance of $\hat{O}_{s,m}^{SC}$ does not matter if one is deep inside the
Mott phase and a large gap in the $\phi_{\rho+}$ field causes strong
fluctuations of $\theta_{\rho+}$, thus suppressing any SC
proximity effect.  This would be the case in a system described in
Ref.\onlinecite{Carr-CNT-dimerization} where the dimerization term $H_{di}$
was governed by the Mott gap. In this work we have found another mechanism where the field $\eta(x)$ is locked by
the symmetry breaking $H_{SO}$, and thanks to that the field $\theta_{\rho+}$ is not randomly fluctuating at the lowest energies. Considering the relevance of the SC-proximity now makes sense.

The presence of $M_{\rho-},M_{\sigma+}$, or to be more precise the field configuration they impose, sets a constraint on the allowed proximity
effect. If we disregard the $\sigma-$ mode for a moment, then we find that there
exists one SC order parameter which is compatible with the locked
fields $\langle\phi_{\rho-}\rangle=0$ and
$\langle\phi_{\sigma+}\rangle=0$. It is the $s-SC$ that is also the most likely candidate from the microscopic viewpoint. This order has a
topologically trivial character. The other $O_{m}$ order parameter is
suppressed because it requires to lock the $\phi_{\sigma+}$ field at
the other minimum: $\phi_{\sigma+}=\pi/2$. The triplet order
parameters are exponentially suppressed because they involve
the $\theta_{\sigma+}$ field which is canonically conjugate to the locked $\phi_{\sigma+}$.

As for the $\sigma-$ mode there are two options:
\begin{itemize}
\item\textbf{the mode stays on a self-duality point:} Then $\cos\theta_{\sigma-}$ has a finite expectation value and s-SC is allowed;
\item\textbf{ultimately the gap opens, at much reduced energies:} this will be most likely a gap in the $\phi_{\sigma-}$ field, $m_{\sigma-}$. It could suppress the s-SC proximity effect at the lowest energies. One way to overcome it is to take a sufficiently large amplitude of the proximity induced gap$\Delta_{s}^{SC}>m_{\sigma-}$. Thanks to a huge difference of energy scales between the different masses the 1D character of the system will be still protected by $M_{\rho-}$. The induced transition to the s-SC state shall have the Ising character\cite{Fabrizio-ionicMott} (one of Ising disordered operators $\mu_1$ acquires a finite value at the cost of the Ising ordered operator $\sigma_1$ )
\end{itemize}
In either case only the topologically trivial s-SC is allowed. The Majorana surface states are never allowed to occur.

\section{Discussion}\label{sec:discus}

\subsection{Size of the spectral gap and the means of its detection}\label{ssec:gap-size}

The estimate for spectral gaps that we have given in Sec.\ref{ssec:low-en} is rather conservative, valid
for a CNT embedded in a good dielectric, for instance a CNT suspended in vacuum. By introducing an extra screening,
for example by placing a tube on a superconducting substrate or
within a multi-wall CNT, one makes electron-electron interactions more local. In reciprocal space this increases the large momentum exchange component of electron-electron interactions, $V(q\approx 2K)$. Then the bare amplitudes of
the backscattering terms $g_i$ in Eq.\ref{eq:Hx-intro} can grow
substantially. Moreover, as we indicate in the context of the proximity effect, placing the tube on an appropriately chosen substrate may introduce additional periodic potentials that cause backscattering and adds up with $g_{d}$ and $g_{so}$. The magnitude of gaps depend on particular experimental realizations and in some circumstances it can be detectable already at energies $\sim 1meV$.

One possibility to detect the $M_{\rho-},M_{\sigma+}$, is to study the Knight shift and relaxation rate of
NMR signal. The temperature dependence shall be a power law but at
the energy scale corresponding to the gap one should observe a
change of an exponent, such an effect was indeed
experimentally\cite{Singer-CNT-NMR1st, Dora-CNT-NMR2nd} observed
but its origin was unclear. In our mechanism, for instance for
the Knight shift we predict a change from
$(K_{\sigma+}+K_{\sigma-})/2$ to $K_{\sigma-}/2$. Moreover, a
known feature of the spin-valley dependent split $V_{so}$ is that
it can be varied by applying an external magnetic
field\cite{minot:nature2004, kuemmeth:nature2008}. Since both
$M_{\rho-}$ and $M_{\sigma+}\sim \tilde{g}_d \sim V_{so}$, and the spin/valley dependent part of the split in a single-particle dispersion can be varied by a magnetic field directed along a tube, then an anisotropic magnetic field dependence of spectral gaps can be taken as a hallmark of
their many-body origin.

\subsection{Relation to SC order parameters proposed for CNTs}\label{ssec:other-SC}

The $\hat{O}_{s}^{SC}$ in the same form like Eq.\ref{eq:s-SCdef}
was also proposed by Egger\cite{Egger-CNT-TLL-big}. The fermionic expression, in the reciprocal space,
for superconducting order parameter that we invoked $\hat{O}_{s}^{SC}$ reads:
$$
\hat{O}_{s}^{SC}=(c^{\dag}_{kK\uparrow}c^{\dag}_{-kK'\downarrow}+h.c.)-(\uparrow\leftrightarrow\downarrow)
$$

and it is equivalent to an inter-chain ordering as derived in a seminal
paper\cite{Khveshchenko-doubl-chain-bos}. In the last paper it is called d-SC, but this should not lead to any misunderstanding, since we
define order parameters for real-space hexagonal lattice, what is
$\hat{O}_d$ for a square ladder is not necessarily d-wave for
other underlying crystal lattice. A detail description of the symmetry properties for a bi-layer graphene interface is given in  Ref.\onlinecite{AnnBS-graph-bi} where a tables of characters for the local
$\hat{O}_{s}^{SC}$, Eq.\ref{eq:s-SCdef} as well as the non-local
$\hat{O}_{m}^{SC}$, the Eq.\ref{eq:m-SCdef}, were found. In particular it was explicitly shown that only the $\hat{O}_{m}^{SC}$ may
contain topologically non-trivial SC order.

Furthermore, one notices that $\hat{O}^{SC}_s$ is different from the
superconducting order parameters proposed previously for the
armchair CNTs\cite{LeHur-CNT-SC}. This is because the band
structure is different: the inter-band order parameter, that was
previously prohibited due to the conservation of $k_{||}$, now is
allowed because in zig-zag (like) tubes the chains of ladder are
associated with valleys and Dirac cones are located at $K_{||}=0$.
Moreover, if the circumferential momentum is conserved, then by
requiring $\vec{k}_1=-\vec{k}_2$ within the BCS pair, we find that
indeed the inter-chain (inter-valley) $O^{SC}_s$ is favored (see Appendix for details). Moreover, from a basic symmetry argument, we know that
the inter-valley Andreev reflection is protected (vs for instance
disorder) by time reversal symmetry. On the other hand the intra-valley pairing would be
protected by the so called symplectic symmetry, but this one is
already broken from the very beginning by introducing the
$\Delta_{SO}$.

So far we have discussed the relation between $\hat{O}_{s,m}^{SC}$ and and other uniform SC-orders proposed before. A novel aspect of proximity effect, that is inevitably present in chiral tubes, is its non-uniformity. For a chiral tube that is rolled along the helical
line one may consider the skew-turn to be a built-in rotation angle
vs substrate lattice. Since strength of bonding is related to interatomic distance,the de Moire pattern of the substrate-tube hybridization appears and the proximity effect is not any longer uniform but instead it becomes periodic\cite{Herman-dM-pattern} (see Appendix for details). Such periodic proximity effect is favourable for more exotic SC orders proposed\cite{Jaefari-non-uniform} for the two leg ladder models and known as pair density waves. One defines a composite order parameter\cite{Jaefari-non-uniform}, that is a product of the $\hat{O}_{s,m}^{SC}$ and some density wave. The density wave shall be defined in the intra-valley channel to avoid a direct competition with superconductivity. One advantage is that one can construct an operator $O_{PDW}^{SC}$ which, albeit less relevant, depends only on $\phi_{\sigma+}$ in the spin sector. This allows to avoid a potential
problem if a field $\phi_{\sigma-}$ is after all locked. The SC order most likely retains a topologically trivial character, in the sense that standard procedure of Ref.\onlinecite{Jaefari-non-uniform} again favours $\hat{O}_{s}^{SC}$. There are many other fascinating aspects of non-uniformity that should stimulate research in this direction. One is that the SC proximity
effect shall be particularly strong for a chemical potential for
which $k_{dM}=k_F$. This opens an exciting perspective of gate
tuning of SC order in CNTs.

\subsection{Other symmetry breaking terms; valley mixing}\label{ssec:deltaKK}

For completeness we comment on other backscattering operators, analogous to $H_{di}$ and $H_{so}$, that
can be introduced into the hamiltonian of a CNT. One frequently
proposed perturbation is an inter-valley backscattering, the so called
$\Delta_{KK'}$. No matter what the content in the spin-space we choose,
this operator written in bosonic language contains
$\theta_{\rho-}$ field, a field canonically conjugate to all
cosine terms present in
Eq's.\ref{eq:Hx-intro},\ref{eq:H-dim},\ref{eq:H-so}. This means
that $\Delta_{KK'}$ is quickly suppressed by all other terms as
one is moving along the RG trajectory (towards $L\rightarrow\infty$ in
quantum dot language). Moreover, even in the case when an
extremely strong $\Delta_{KK'}$ is able to dominate the physics, since the pairing operators in
Eq's.\ref{eq:s-SCdef},\ref{eq:m-SCdef} contain $\cos\phi_{\rho-}$
then none of these (including the potentially topologically non-trivial,
non-local $\hat{O}_{m}^{SC}$) shall be favored. It seems that the $\Delta_{KK'}$ reduces the propensity of the system to any standard proximity effect.

One can also ask an opposite question: what symmetry breaking term could potentially support the $\hat{O}_{m}$ order? A brief inspection of all order parameters reveals that this is a rather exotic spiral
electric field acting opposite on two valleys (but
valley-diagonal), however this would need to be taken together with attractive
$V(q\approx 2|K|)$ interactions.

\section{Conclusions}\label{sec:concl}

We have shown that, for a chosen sub-set of CNTs, the
presence of spin-orbit coupling $\Delta_{so}$ leads to a gap
opening in the spectrum of two bosonic modes $\phi_{\rho-}$ and
$\phi_{\sigma+}$ . This drastically reduces the sub-set of
proximity effects allowed at the lowest energies: we find that
only a phase with a trivial topology is allowed. This statement is
quite general as it should remain valid also upon increasing the
interaction strength, doping, hybridization with the substrate
and upon adding another symmetry breaking term $\Delta_{KK'}$.
An extra motivation, and a broader perspective, for this work
comes from the recently synthesized 2D analogs of graphene: silicene,
germacene and stanene. This gives a hope for a new class of
nanotubes that shall be built out of atoms heavier than carbon.
Since $\Delta_{SO}\sim \lambda_{so}$ (where $\lambda_{so}$ is an
atomic spin-orbit coupling constant\cite{Jeong-curvature-SO}),
then the fine effects predicted here can become orders of
magnitude larger.

\appendix

\section{Estimate for $\langle\cos\tilde{\phi}_{\rho+}\rangle|_{\omega=\Lambda'}$}\label{app:holon-lambdap}

The dynamics of the fast field $\phi_{\rho+}(x)$ for energies $\sim \Lambda'$ is rather
complicated. Usually $\delta \sim 1$ implies
that the  $g_d$ and $g_{so}$ terms rapidly drop out of the problem
(and $g_3$ as well, but by writing Eq.\ref{eq:ham-fast} we had
already neglected $g_3$). However, already in the simplest single mode sine Gordon model, the issue of how precisely the expectation values disappear when $\delta$ becomes substantial, has proven to be quite-nontrivial and depends on how precisely RG procedure is set up\cite{Horovitz-RGcutoff}. Our model is much more complicated as the dynamic coupling with three other modes is present. For instance in the argument of the cosine one can clearly see the competition between $\eta(x)$ and $\delta x$. Moreover, the amplitude of the cosine, that is $M[\phi_{i\neq \rho+}]$, shall have an extra increase when the two neutral fields order.

To tackle the problem let us assume that, to begin with when $\eta(x)\approx 1$, $\eta(x)$ dominates. At $\Lambda'$ energy scale the neutral fields still
fluctuate, with a velocity that is irrational with the holon velocity, thus $M[\phi_{i\neq \rho+}]$ and the $\eta(x)$ field can be
considered as amplitude and phase of a complex random variable.
Then Eq.\ref{eq:ham-fast} can be interpreted as a model of a
random backscattering in a TLL. The suppression of $g_d$ is delayed by the
fact that for the disorder problem the scaling dimension is even
larger $d_{dis}=3-2K_{\rho+}$. One can say that a strong enough disorder freezes the correlation function for $l\approx\Lambda'$. Such a phenomenon occurs also for incommensurate (not necessarily random) potentials when modes of different velocities couple. It is then known as Aubry-Andr{\'e} transition\cite{aubry1980analyticity}. The
crossover is quite complex, but most likely as the energy scale $l$
decreases during the RG, then $\eta \rightarrow 0$ (the
fluctuations cease below the energy $\sim M_{\rho-}$), the
randomness disappears and $g_{3,d,so}[l]$ resume their flow to
zero, driven by a finite doping. However, for energies
around $\Lambda'$, the sine-Gordon model, Eq.\ref{eq:ham-fast},
with a finite, energy independent amplitude of the cosine term, gives a correct
description. Then one can attempt to compute
$\langle\cos\tilde{\phi}_{\rho+}\rangle|_{\omega=\Lambda'}$, in a limit when the shift goes to zero, by using
results known from the Ising model in the renormalized classical regime.
One may either use the zero temperature result $\sim K_0(\tilde{g}_{1c}\tau')$,
where $K_0$ is the modified Bessel function of the second kind and a characteristic time-scale is set as $\tau'\sim1/\Lambda'$, or a finite
temperature result Ref.\onlinecite{Sachdev-corr} where $\langle\cos\phi_{\rho+}\rangle|_{\omega=\Lambda'}$
is proportional to $erfc(\sqrt{\tilde{T}/2\tilde{g}_{1c}})$ with
a characteristic temperature taken to be $k_B \tilde{T} = \Lambda'$ (and $erfc$ is a complementary error function). In both
estimates we get
$\langle\cos\phi_{\rho+}\rangle|_{\omega=\Lambda'}\sim 10^{-1}$. We consider it as an upper limit for $\langle\cos\tilde{\phi}_{\rho+}\rangle|_{\omega=\Lambda'}$ and in all further calculations we take a more conservative value $\langle\cos\tilde{\phi}_{\rho+}\rangle|_{\omega=\Lambda'}= 10^{-2}$.


\section{Details of an overlap with a substrate}\label{app:overlap-proxim}

In Ref.\onlinecite{Affleck-proximity-basic} it is shown that every
site which is in touch with a superconductor, upon integrating out
the BCS condensate, acquires an emergent pairing potential:$
\int dx dy \Delta_{sc}(x,y)(c^{\dag}_{\bar{\sigma}}(x,y)c^{\dag}_{-\bar{\sigma}}(x,y)+h.c.)$
where $(x,y)$ are the site coordinates (interface has 2D character) and $\Delta_{sc}(x,y)\sim t"(x,y)^2/V_F$ is a pairing strength, with $t"(x)$
a hybridization between CNT and a substrate. Let us consider a process of creation of a Cooper
pair inside a CNT:
$c^{\dag}_{\vec{k_1}}c^{\dag}_{\vec{k_2}}$. For a moment we need to take a 2D $\vec{k_1}$ because we keep interfaces' 2D character. When the pair
is created in two different valleys (an inter-valley term), it is compatible with the
standard s-wave BCS pairs in the substrate where $k_1=-k_2$. Contrarily, the intra-valley term
does not conserve momenta since then $k_{1\perp}=k_{2\perp}\pm
2K_{\perp}$. Thus this second process will be suppressed when $k_{i\perp}$ is a conserved quantity during the tunneling process. We can
try to quantify the condition for conservation of the circumferential momenta. We take the hybridization $t"(x, y)$ to be a Gaussian with a width
proportional to the nanotube radius: $\delta y = \alpha_b R$, where $\alpha_b$ is some proportionality constant and $y$ is a direction along tube's circumference. This relation simply encodes the fact that for broader tubes there are more carbon atoms that can build a covalent bond with a substrate. Finite $\delta y$ produces
a momentum resolution $\delta k_{perp}\approx 1/\delta y $. The
two valleys can be distinguished provided the real-space Gaussian
is broad enough, that is $\delta k_{perp}<|K|\Leftrightarrow
R>a/|K|$. When this condition is fulfilled one can consider valley
index and thus $k_{\perp}$ to be a conserved quantity in a
substrate-CNT tunneling process.

The microscopic model also allows us to take a closer look at the non-uniformity of the $t"(x)$.
For a chiral tube the hexagonal lattice makes consecutive
skew-turns around the central axis of the tube. Then looking from
the top it is very much like a sequence of tilted hexagons ($\delta y>\sqrt{3}a$, with $a$ graphene lattice constant, is assumed). If one
puts two hexagonal lattices one on the top of another and rotate
(or re-scale) one of them then he obtains the periodic de Moire
pattern. Re-scaling is necessary only when the substrate is a crystal different from graphene. We conclude that a chiral CNT placed on the top of a 2D surface,
gives an effective hybridization $t"(x)=\frac{1}{\delta y}\int dy t"(x,y)$ that is not constant along the
tube but varies and these variations are the strongest for smaller tubes where the effect is not averaged out by integration over large $\delta y$. For scaling factor between two lattices equal
to one (e.g. both based on graphene) one finds\cite{Herman-dM-pattern} that the angle between the
$\vec{k}_{dM}$ and the CNT basis is $\pi/2$ and indeed the
hybridization $t"(x)$ along 1D profile is periodic. The patterns'
periodicity depend on the chiral angle, for small chiral angles
very small $|\vec{k}_{dM}|$ can be reached.

\bibliography{CNTbiblio}

\begin{thebibliography}{10}

\bibitem{Kane-topoSC}
Liang Fu and C.~L. Kane.
\newblock Superconducting proximity effect and majorana fermions at the surface
  of a topological insulator.
\newblock {\em Phys. Rev. Lett.}, 100:096407, Mar 2008.

\bibitem{Fujimoto-topoSC2}
Satoshi Fujimoto.
\newblock Topological order and non-abelian statistics in noncentrosymmetric
  $s$-wave superconductors.
\newblock {\em Phys. Rev. B}, 77:220501, Jun 2008.

\bibitem{Wilczek-viewpoint-Majorana}
Frank Wilczek.
\newblock Majorana returns.
\newblock {\em Nat.Phys.}, 5:614, 2009.

\bibitem{Alicea-topSC-prox}
Jason Alicea.
\newblock Majorana fermions in a tunable semiconductor device.
\newblock {\em Phys. Rev. B}, 81:125318, Mar 2010.

\bibitem{Kouwenhoven-the-Maj-experim}
V.~Mourik, K.~Zuo, S.~M. Frolov, S.~R. Plissard, E.~P. A.~M. Bakkers, and L.~P.
  Kouwenhoven.
\newblock Signatures of majorana fermions in hybrid
  superconductor-semiconductor nanowire devices.
\newblock {\em Science}, 336(6084):1003--1007, 2012.

\bibitem{Atland-disord}
Dmitry Bagrets and Alexander Altland.
\newblock Class $d$ spectral peak in majorana quantum wires.
\newblock {\em Phys. Rev. Lett.}, 109:227005, Nov 2012.

\bibitem{Lee-disord}
Jie Liu, Andrew~C. Potter, K.~T. Law, and Patrick~A. Lee.
\newblock Zero-bias peaks in the tunneling conductance of spin-orbit-coupled
  superconducting wires with and without majorana end-states.
\newblock {\em Phys. Rev. Lett.}, 109:267002, Dec 2012.

\bibitem{Braunecker-interac}
Suhas Gangadharaiah, Bernd Braunecker, Pascal Simon, and Daniel Loss.
\newblock Majorana edge states in interacting one-dimensional systems.
\newblock {\em Phys. Rev. Lett.}, 107:036801, Jul 2011.

\bibitem{Kane+Mele-gaps}
C.~L. Kane and E.~J. Mele.
\newblock Size, shape, and low energy electronic structure of carbon nanotubes.
\newblock {\em Phys. Rev. Lett.}, 78:1932--1935, Mar 1997.

\bibitem{ando:jpsj2000}
T.~Ando.
\newblock Spin-orbit interaction in carbon nanotubes.
\newblock {\em J. Phys. Soc. Jpn}, 69:1757, 2000.

\bibitem{Jeong-curvature-SO}
Jae-Seung Jeong and Hyun-Woo Lee.
\newblock Curvature-enhanced spin-orbit coupling in a carbon nanotube.
\newblock {\em Phys. Rev. B}, 80:075409, Aug 2009.

\bibitem{egger:prb2012}
R.~Egger and K.~Flensberg.
\newblock Emerging dirac and majorana fermions for carbon nanotubes with
  proximity-induced pairing and spiral magnetic field.
\newblock {\em Phys. Rev. B}, 85:235462, 2012.

\bibitem{sau:prb2013}
J.D. Sau and S.~Tewari.
\newblock Topological superconducting state and majorana fermions in carbon
  nanotubes.
\newblock {\em Phys. Rev. B}, 88:054503, 2013.

\bibitem{KaneBalents-armchair}
Charles Kane, Leon Balents, and Matthew P.~A. Fisher.
\newblock Coulomb interactions and mesoscopic effects in carbon nanotubes.
\newblock {\em Phys. Rev. Lett.}, 79:5086--5089, Dec 1997.

\bibitem{Egger-CNT-TLL-big}
R.~Egger and A.~O. Gogolin.
\newblock Correlated transport and non-fermi-liquid behavior in single-wall
  carbon nanotubes.
\newblock {\em Eur. Phys. J. B}, 3(3):281, 1998.

\bibitem{kuemmeth:nature2008}
F.~Kuemmeth, S.~Ilani, D.C. Ralph, and P.L. McEuen.
\newblock Coupling of spin and orbital motion of electrons in carbon nanotubes.
\newblock {\em Nature}, 452:448, 2008.

\bibitem{jespersen:prl2011}
T.S. Jespersen, K.~Grove-Rasmussen, K.~Flensberg, J.~Paaske, K.~Muraki,
  T.~Fujisawa, and J.~Nyg{\aa}rd.
\newblock Gate-dependent orbital magnetic moments in carbon nanotubes.
\newblock {\em prl}, 107:186802, 2011.

\bibitem{Desphande-Nature-rev}
Vikram~V. Deshpande, Marc Bockrath, Leonid~I. Glazman, and Amir Yacoby.
\newblock Electron liquids and solids in one dimension.
\newblock {\em Nature}, 464:209, 2010.

\bibitem{Egger-diagonal-spin-orbit}
Andreas Schulz, Alessandro De~Martino, and Reinhold Egger.
\newblock Spin-orbit coupling and spectral function of interacting electrons in
  carbon nanotubes.
\newblock {\em Phys. Rev. B}, 82:033407, Jul 2010.

\bibitem{Carr-CNT-dimerization}
Sam~T. Carr, Alexander~O. Gogolin, and Alexander~A. Nersesyan.
\newblock Interaction induced dimerization in zigzag single wall carbon
  nanotubes.
\newblock {\em Phys. Rev. B}, 76:245121, Dec 2007.

\bibitem{Eggert-gap-twist}
Alex Kleiner and Sebastian Eggert.
\newblock Band gaps of primary metallic carbon nanotubes.
\newblock {\em Phys. Rev. B}, 63:073408, Jan 2001.

\bibitem{moj-2classCNT}
Magdalena Marganska, Piotr Chudzinski, and Milena Grifoni.
\newblock Coulomb interactions and mesoscopic effects in carbon nanotubes.
\newblock {\em arxiv.org/cond-mat}, 1412.7484, Dec 2014.

\bibitem{Tsuchiizu-commen-incommen}
Tsuchiizu M., Donohue P., Suzumura Y., and Giamarchi T.
\newblock Commensurate-incommensurate transition in two-coupled chains of
  nearly half-filled electrons.
\newblock {\em The European Physical Journal B - Condensed Matter and Complex
  Systems}, 19(2):185--193, 2001.

\bibitem{moj-ladder}
P.~Chudzinski, M.~Gabay, and T.~Giamarchi.
\newblock Orbital current patterns in doped two-leg cu-o hubbard ladders.
\newblock {\em Phys. Rev. B}, 78:075124, Aug 2008.

\bibitem{Nersesyan-CNT-orders}
A.~A. Nersesyan and A.~M. Tsvelik.
\newblock Coulomb blockade regime of a single-wall carbon nanotube.
\newblock {\em Phys. Rev. B}, 68:235419, Dec 2003.

\bibitem{Levitov-narrow-gaps-TLL}
L.~S. Levitov and A.~M. Tsvelik.
\newblock Narrow-gap luttinger liquid in carbon nanotubes.
\newblock {\em Phys. Rev. Lett.}, 90:016401, Jan 2003.

\bibitem{minot:nature2004}
E.~D. Minot, Y.~Yaish, V.~Sazonova, and P.~McEuen.
\newblock Determination of electron orbital magnetic moments in carbon
  nanotubes.
\newblock {\em Nature}, 428:536, 2004.

\bibitem{gogol-tsvel_book_1d}
Gogolin A.O., Nersesyan A.A., and Tsvelik A.M.
\newblock {\em Bosonization and Strongly Correlated Systems}.
\newblock Cambridge University Press, Cambridge, 1998.

\bibitem{Fabrizio-ionicMott}
Michele Fabrizio, Alexander~O. Gogolin, and Alexander~A. Nersesyan.
\newblock From band insulator to mott insulator in one dimension.
\newblock {\em Phys. Rev. Lett.}, 83:2014--2017, Sep 1999.

\bibitem{Singer-CNT-NMR1st}
P.~M. Singer, P.~Wzietek, H.~Alloul, F.~Simon, and H.~Kuzmany.
\newblock Nmr evidence for gapped spin excitations in metallic carbon
  nanotubes.
\newblock {\em Phys. Rev. Lett.}, 95:236403, Nov 2005.

\bibitem{Dora-CNT-NMR2nd}
Bal\'azs D\'ora, Mikl\'os Gul\'acsi, Ferenc Simon, and Hans Kuzmany.
\newblock Spin gap and luttinger liquid description of the nmr relaxation in
  carbon nanotubes.
\newblock {\em Phys. Rev. Lett.}, 99:166402, Oct 2007.

\bibitem{Khveshchenko-doubl-chain-bos}
D.~V. Khveshchenko and T.~M. Rice.
\newblock Spin-gap fixed points in the double-chain problem.
\newblock {\em Phys. Rev. B}, 50:252--257, Jul 1994.

\bibitem{AnnBS-graph-bi}
Annica~M. Black-Schaffer and Sebastian Doniach.
\newblock Resonating valence bonds and mean-field $d$-wave superconductivity in
  graphite.
\newblock {\em Phys. Rev. B}, 75:134512, Apr 2007.

\bibitem{LeHur-CNT-SC}
Karyn Le~Hur, Smitha Vishveshwara, and Cristina Bena.
\newblock Double-gap superconducting proximity effect in armchair carbon
  nanotubes.
\newblock {\em Phys. Rev. B}, 77:041406, Jan 2008.

\bibitem{Herman-dM-pattern}
Klaus Hermann.
\newblock Periodic overlayers and moiré patterns: theoretical studies of
  geometric properties.
\newblock {\em Journal of Physics: Condensed Matter}, 24(31):314210, 2012.

\bibitem{Jaefari-non-uniform}
Akbar Jaefari and Eduardo Fradkin.
\newblock Pair-density-wave superconducting order in two-leg ladders.
\newblock {\em Phys. Rev. B}, 85:035104, Jan 2012.

\bibitem{Horovitz-RGcutoff}
B.~Horovitz, T.~Bohr, J.~M. Kosterlitz, and H.~J. Schulz.
\newblock Commensurate-incommensurate transitions and a floating devil's
  staircase.
\newblock {\em Phys. Rev. B}, 28:6596--6599, Dec 1983.

\bibitem{aubry1980analyticity}
Serge Aubry and Gilles Andr{\'e}.
\newblock Analyticity breaking and anderson localization in incommensurate
  lattices.
\newblock {\em Ann. Israel Phys. Soc}, 3(133):18, 1980.

\bibitem{Sachdev-corr}
Subir Sachdev.
\newblock Universal, finite-temperature, crossover functions of the quantum
  transition in the ising chain in a transverse field.
\newblock {\em Nuclear Physics B}, 464(3):576 -- 595, 1996.

\bibitem{Affleck-proximity-basic}
Ian Affleck, Jean-S\'ebastien Caux, and Alexandre~M. Zagoskin.
\newblock Andreev scattering and josephson current in a one-dimensional
  electron liquid.
\newblock {\em Phys. Rev. B}, 62:1433--1445, Jul 2000.

\end{thebibliography}

\end{document}